# 3D transcranial Dynamic Ultrasound Localization Microscopy in the mouse brain using a Row-Column Array

Alice Wu[1], Jonathan Porée[1], Gerardo Ramos-Palacios[2], Chloé Bourquin[1], Nin Ghigo[1], Alexis Leconte[1], Paul Xing[1], Abbas F. Sadikot[2], Michaël Chassé[3] and Jean Provost[1,4]

[1]Department of Engineering Physics, Polytechnique Montréal, Montréal, QC H3T 1J4, Canada
[2]Montreal Neurological Institute, McGill University, Montréal, QC H3A 2B4, Canada
[3]Department of Medicine, Université de Montréal, Montreal, QC, Canada
[4]Montreal Heart Institute, Montréal, QC H1T 1C8, Canada

*Abstract* — The role of brain hemodynamics in neurodegenerative diseases cannot be fully assessed using existing imaging technologies. Recently, 2D Dynamic Ultrasound Localization Microscopy (DULM) has allowed for the quantitative mapping of the pulsatile flow at sub-wavelength resolution. However, to obtain accurate velocity estimates, 3D imaging is more adapted, especially for complex vascularized organs like the brain. 3D+t DULM is achievable using matrix array probes, but suffers from limitations in terms of cost, device complexity associated with the high channel count, and operating frequencies. Alternatively, Row Column Arrays (RCA) can reduce the number of elements while maintaining a large field of view and high frame rate. Herein, we demonstrate the feasibility of performing 3D+t blood flow measurements in the mouse brain using an RCA and a DULM sequence with a high spatiotemporal resolution. Transcranial images of anesthetized mice (n=7) were acquired at a volume rate of 750 Hz using 42 tilted plane waves. After microbubbles localization and tracking, super-resolved dynamic density and velocity maps of the 3D brain vascular network were obtained. Cortical vessels were segmented and pulsatility in the arteries was significantly higher than in veins for all mice, in accordance with the literature. Our results demonstrate the feasibility and reproducibility of achieving high spatiotemporal resolution volumes of the mouse brain vasculature with DULM using a RCA.

*Index Terms* — Cerebral blood flow, Dynamic Ultrasound Localization Microscopy, Pulsatility Index, Row-Column Arrays, Super-resolution imaging, Transcranial brain imaging, Volumetric imaging.

## I. INTRODUCTION

Brain hemodynamics dysfunction correlates with cognitive impairment and neurodegenerative diseases [1], [2]. A potential marker in this context is the pulsatility index (PI) in cerebral arteries, which measures the relative variation of flow throughout the cardiac cycle. Specifically, Alzheimer's disease (AD) patients have a higher PI in cerebral arteries compared to nondemented subjects [3]. Additionally, elevated pulsatility levels, even in individuals without dementia, indicate potential cognitive decline in subsequent years [4]. This highlights the importance of accurate pulsatility measurements in cognitive health research. Recent studies showed that cerebral vascular system is significantly disrupted in the Huntington's Disease (HD) [5], [6], a genetic disorder that causes progressive degeneration of nerve cells in the brain, leading to motor dysfunction, cognitive decline, and psychiatric symptoms. Understanding the vascular changes in both AD and HD highlights the broader significance of cerebral hemodynamics in neurodegenerative conditions.

Optical techniques such as optical coherence tomography (OCT) [7], [8] or two-photon microscopy (TPM) [9] provide spatiotemporally resolved images of the brain down to capillaries but are limited to depths of few millimeters. Even though Transcranial Doppler (TCD) ultrasound was proposed as a tool to monitor neurodegenerative disease [10], [11], it remained limited in assessing blood flow pulsatility in major cerebral arteries due to its limited sensitivity and spatial resolution.

Ultrasound Localization Microscopy (ULM) [12], [13], [14] enables microvascular imaging down to a 10-µm resolution while requiring a few minutes of acquisition [15]. ULM is based on sub-wavelength localization and tracking of microbubbles (MBs) injected into the bloodstream and has been applied to the imaging of brain microvasculature in animal models, both in 2D [12], [16], [17] and 3D [18], [19], [20], [21], and in human in 2D [22]. Dynamic Ultrasound Localization Microscopy (DULM) [23], [24] has been recently proposed to extend the capabilities of ULM by adding a temporal dimension using retrospective gating [23], [25]. 2D DULM can be used to map brain-wide pulsatility in microvessels [23], [26]. 2D fields of view, however, limit studies due to intra- and inter-observer variability.

Several approaches have been suggested to expand ULM for

This work was supported in part by the Institute for Data Valorization (IVADO), in part by the Canada Foundation for Innovation under Grant 38095, in part by the Canadian Institutes of Health Research (CIHR) under Grant 452530, in part by the New Frontiers in Research Fund under Grant NFRFE-2018-01312 and in part by the Natural Sciences and Engineering Research Council of Canada (NSERC) under Grant RGPIN-2019-04982. Further support came from the TransMedTech Institute, the Fonds de recherche du Québec - Nature et technologies, the Quebec Bio-Imaging Network, the CONAHCYT, the Insightec, the Healthy Brains Healthy Lives, the Canada First Research Excellence Fund, and the NSERC. Additionally, computational resources were provided through the Digital Research Alliance of Canada. (Corresponding author: Jean Provost).

3D imaging such as fully-addressed or multiplexed 2D matrix-array transducers [18], [19], [20], [27], [28], 1.5-D arrays [29], sparse 2-D arrays [30], [31], [32], and Row Columns Arrays (RCA) [33], [34], [35], [36], [37], [38], [39]. Fully addressed matrix-array probes require a large number of small piezoelectric elements and sampling channels, leading to limited sensitivity, high manufacturing cost, and complexity. Multiplexing matrix arrays can partially lower the cost but at the expense of volume rate [19]. Sparse arrays reduce the number of elements but at the cost of higher side lobe levels and thus lower sensitivity. RCA, which are equivalent to two linear probes with large elevation orthogonally overlapped, reduce the number of elements and channels while keeping a large field of view and a relatively high-volume rate (750 Hz). RCA have been used for 3D functional imaging of the rat brain [40] and super resolution of the rat kidney [41] and rabbit kidney [42], and human thyroid [42]. 3D DULM has been recently developed using a multiplexed matrix probe [26], but the reduced imaged quality limited quantitative pulsatility measurements to larger vessels.

In this study, we show the feasibility and reproducibility of RCA probes performing 3D non-invasive transcranial DULM in the mouse brain (n=7) and achieving spatiotemporally resolved dynamic vascular volumes. Pulsatile blood flow and pulsatility indices were obtained in both small and large vessels of the brain (from 50 μm to 200 μm diameter) at various depths. We establish the feasibility of 3D DULM in young R6/2 mouse model of HD [43]. We show a significant difference in pulsatility indices between arteries and veins in the cortex of all mice studied (n=7, wild-type C57BL/6 and R6/2) along with comparable velocity distributions, indicating reproducibility. Thus, 3D transcranial DULM using RCA can produce highly resolved dynamic volumes of the blood flow non-invasively, providing a robust, less complex, and less costly setup to better understand brain hemodynamics.

II. METHODS

*A. Ethics*

This study was carried out in accordance with the recommendations of the guide for the care and use of laboratory animals of the Canadian Council for Animal Care. All procedures were approved by the Neuro Animal Care Committee and McGill University (protocol number 4532).

*B. Animal Preparation*

Animals were anesthetized under 2% isoflurane while the body temperature was maintained at 37°C with a water circulating heating pad. The head of the animal was placed in a stereotaxic frame to limit motion after removing hair (skin and skull kept intact). One milliliter of warm saline solution was injected in intraperitoneal to keep the animal hydrated during anesthesia. Acoustic coupling was ensured by centrifuged ultrasound gel on the head of the animal. A 3D printed water tank was then placed above the gel with a transparent latex membrane window for the positioning of the probe and to ensure appropriate ultrasound coupling as the aperture of the probe is larger than the mouse head.

The ultrasound acquisitions were launched after an intravenous bolus injection in the tail vein of a microbubble (MB) solution at 4 μL/g of bodyweight ($1.2 \times 10^{10}$ MBs per milliliter, Definity, Lantheus Medical Imaging, MA, USA, diluted in sterile phosphate-buffered saline (PBS) at 7.4 pH in a different ratio immediately before injecting) (Fig.1a) with a 30G needle.

*1) Microbubble optimization on one mouse*

One series of acquisitions was first carried out in one mouse (female, 6-week-old, wild-type C57BL/6J, labeled as mouse #1) to optimize the acquisition parameters (sampling, saturation, TGC, voltage, frequency) and MB concentration (Definity:PBS ratio of 1:5, 1:10 and 1:20).

*2) Pulsatility study*

Then, experiments to study pulsatility imaging with the optimized imaging sequence (described below) and MB concentration (Definity:PBS ratio of 1:20) were performed using young female carrier (R6/2, n=3, 4 to 5-week-old) and noncarrier (WT, n=3, 5 to 6-week-old) mice from the same breeding colony (B6CBATg(HDexon1)62Gpb/3J), for further pathological studies. The R6/2 model carries an expanded CAG repeat in exon 1 of the Huntingtin gene and has been shown to have early evidence of blood-brain barrier impairment and cerebral blood flow changes [43], [44]. All mice were 4 to 6 weeks old. Note that the objective of this work is not to study the R6/2 model per se, but rather to develop a methodology that enables 3D DULM in both wild-types and carriers, the latter being imaged at a young age to limit potential vascular differences.

*C. 3D Ultrafast Ultrasound Acquisitions*

In vivo mouse brains were imaged transcranially with a (128+128)-element row-column array probe (including 4 dead elements: elements 72, 73, 93 and 94 in the x subarray) centered at 12 MHz with a pitch of 120 μm (Daxsonics, NS, Canada). The probe was connected to a programmable ultrafast ultrasound system (Vantage 256, Verasonics, WA, USA). $N_{buff}$ buffers of $N_{frame}$ volumes were acquired during 5 (200 buffers) to 7 minutes (240 buffers) at a 750 Hz volume rate and 32 kHz PRF (Fig.1b), using 21 tilted plane waves on each subarray (−5 to 5 degrees). To reject grating lobes, the angular pitch was set according to $\Delta\alpha = \frac{\lambda}{L_{x/y}}$ [40], where, $\Delta\alpha$ is the angular pitch between plane waves in radians, $\lambda$ is the wavelength at the transmitted central frequency $f_c$ and $L_{x/y}$ the aperture of the probe in the direction x and y. Given the aperture of 15.36 mm, an angular pitch of 0.5 degree was selected. As the probe was larger than a mouse head, transmit apodization with a Tukey window with a cosine fraction of 0.6 was applied to limit the transmission field to the brain. We acquired between 1 minute of data for the optimization sequence (200 buffers of 190 frames) and 2 minutes of data for pulsatility study (200-240 buffers of 380 frames). The number of frames $N_{frame}$ per buffer is optimized for pulsatility study mice (R6/2 and WT) to have several cardiac cycles of the animal. A 4-cycle pulse was transmitted at a frequency of 11 MHz at 50 V, corresponding to a mechanical index of 0.13 and a maximum derated spatial peak temporal average intensity (ISPTA) of 61.2 mW/cm$^2$ considering the full aperture of the probe at 8.75 mm depth in a water tank.

### D. 3D ULM Processing

Radio Frequency (RF) data acquired using rows and columns were filtered separately with a Singular Value Decomposition (SVD) (Fig.1c) [45]. In-phase/Quadrature (IQ) beamforming was performed using a delay-and-sum algorithm with the first time arrival of the wavefront to a long element. Orthogonal Plane Wave (OPW) [46] compounding was used to coherently sum the contributions from each subarray to reconstruct volumes with a λ/2 × λ/2 × λ/2 voxel size (Fig.1c) (λ = 128 μm).

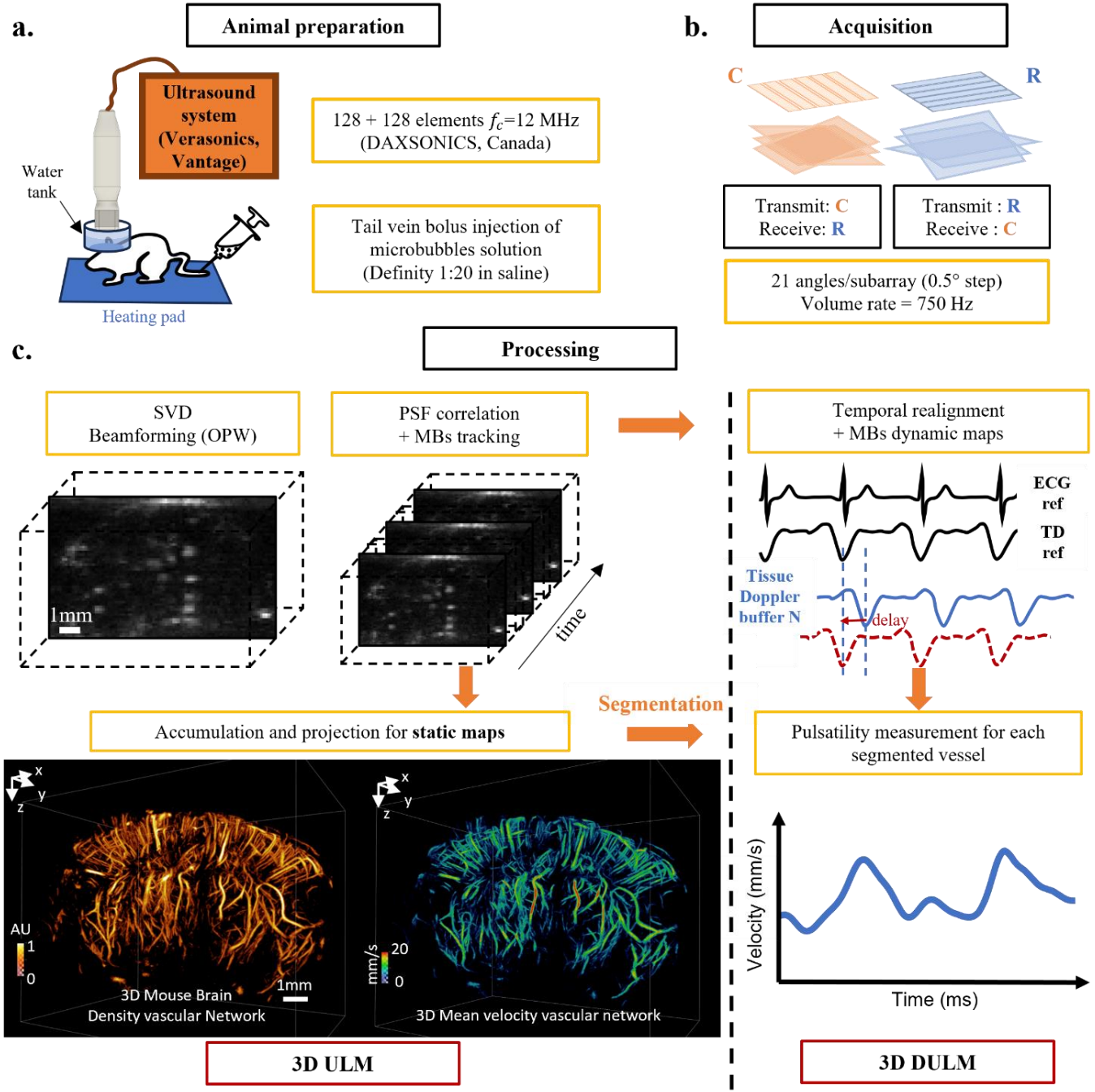

*Figure 1. 3D transcranial DULM experimental set up and processing pipeline using RCA. (a) Animal preparation: mice were anesthetized under isoflurane while the body temperature was maintained at 37°C with a heating pad. Acoustic coupling was ensured by centrifuged ultrasound gel and a 3D printed water tank placed above the gel with a transparent latex for RCA positioning. A bolus of diluted microbubbles solution was injected in the tail vein of the animal before performing (b) the acquisition with 21 angles in each subarray at a volume rate of 750 Hz. (c) After SVD clutter filtering and beamforming with OPW, IQ volumes of flowing MBs were obtained. ULM pipeline starts with MBs localization with a PSF correlation and a gaussian fitting, then tracking using Hungarian algorithm. Accumulation and projection of all MBs in time provide static density and velocity maps (maximum intensity projection). DULM pipeline adds temporal realignment performed with Tissue Doppler (TD) as described in Ghigo et al. (2023) [40], to extract the MBs velocity variation (pulsatility measurements) along the cardiac cycle for each segmented vessel from the density map.*

After manually segmenting the brain from a B-mode volume, IQ volumes were correlated with the Point Spread Function (PSF) simulated using the Verasonics simulator. MBs were detected as local maxima on the IQ envelope volumes and subpixel positions were obtained using a 3D Gaussian fitting with a kernel size of 9 pixels (4.5 λ). To limit false MB detections, bright local maxima with a correlation coefficient lower than 0.3 (optimization acquisitions) or 0.4 (pulsatility sequence) were rejected. Tracking was based on the Hungarian method [47] with a maximum linking distance of one voxel between two consecutives frames corresponding to a maximal velocity of 50 mm/s, and without gap filling (https://github.com/tinevez/simpletracker, Jean-Yves Tinevez, 2021). Only tracks longer than 15 frames (i.e., 20 ms) were kept for pulsatility measurements.

Parallel processing of each buffer was using MATLAB (R2021a, The MathWorks, Inc., Natick, MA) on a cluster computing system with allocated dedicated computational resources provided by the Digital Research Alliance of Canada (NVidia V100SXM2 16G, Intel Gold 6148 Skylake @ 2.4 GHz , 64G RAM).

*E. 3D DULM Processing*

A motion matching approach [48] based on Tissue Doppler (TD) [25] was used to temporally realign tracks and to reject buffers obtained during breathing. Realigned tracks were then smoothed using cubic splines (*csaps* Matlab function with a smoothing factor of 0.8) and velocities were obtained from the closed-form derivative of the spline curve. Static density and velocity volumes were obtained by accumulating and projecting detections on a volume with 20 μm × 20 μm × 20 μm voxel size. Dynamic density and velocity volumes had the same voxel size and a 7-ms temporal resolution during 2 to 3 cardiac cycles.

*F. Pulsatility Measurement in a Vessel*

Vessels were segmented by applying a 3D Hessian filter onto the static density volume followed by skeletonization (*bwskel* Matlab function). MBs velocities were spatially averaged within segmented vessels to produce a vessel-specific temporal velocity variation onto which a gaussian-weighted moving average of 30 frames (i.e., 40 ms) was applied.

The Pulsatility Index (PI) was then calculated as:
$$PI = \frac{PSV - EDV}{MFV} \quad (1)$$
Where $PSV$ is the mean peak systolic velocity, $EDV$ the mean end diastolic velocity, and $MFV$ the mean flow velocity. The PI was calculated over several complete velocity waveforms (2 for the optimization sequence since it is shorter, and 3 for the pulsatility study).

*G. Arteries and veins segmentation in the Cortex*

MB tracks were separated into two groups: flowing upwards and flowing downwards according to the initial and final positions of the MBs in the track. In the cortex, vessels going downwards correspond to the arteries, and those flowing upwards are the veins. PI were calculated for each cortical vessel as described above by manually segmenting the cortex.

*H. Diameter estimation of the segmented vessels.*

For each segmented vessel longer than 10 voxels (centerline length), diameter was estimated as the maximum diameter along the centerline of the vessel using the Euclidian Distance Transform on an interpolated binarized density volume of 10 × 10× 10 μm voxel size.

*I. Spatial resolution*

The spatial resolution was estimated using the Fourier Shell Correlation (FSC) [49]. Briefly, two sub-volumes were obtained by projection as explained in II.E where all the buffers are randomly separated into two datasets. The FSC was computed by using correlation between the spatial spectra $F_1$ and $F_2$ of each sub-volume for voxels within a radius r:
$$FSC(r) = \frac{\sum_{r \in R} F_1(r) \cdot F_2(r)^*}{\sqrt{\sum_{r \in R} |F_1(r)|^2 \cdot \sum_{r \in R} |F_2(r)|^2}} \quad (2)$$
The intersection of the FSC curve with the half-bit threshold was used to establish the resolution.

*J. Statistical analysis*

Statistical analysis was performed using MATLAB. The PI normality distribution of each group of vessels was tested with one-sample Kolmogorov-Smirnov. Then, statistical comparison between cortical veins and arteries was performed using an unpaired two samples Student t-test with unequal variance. A p-value of < 0.05 was considered statically significant. Levels of significance are given as *: p< 0.05, **: p < 0.01, and ***: p< 0.001.

*K. 3D Static and Dynamic Rendering*

Static and dynamic volumes were rendered using the Amira software (2021 version, Thermo Fisher) (see Supplementary videos). 3D density maps were displayed after compressing the values using a 1/3 exponent to better visualize smaller vessels where fewer MBs were detected. No compression factor was applied for velocity maps. Dynamic videos were obtained after a temporal gaussian smoothing of the dynamic volumes (window length of 10 frames). Voxels containing fewer than 2 MBs were set to 0.

III. RESULTS

*A. Effect of MBs Concentration on Image Quality*

3D density volumes for different concentrations of MBs diluted with PBS are shown in Fig.2a and 1-mm thick coronal slices of the mouse brain (around Bregma - 1.6 mm and - 3.0 mm respectively) are shown in Fig.2b. From 1:5 to 1:20 concentration, smaller vessels are reconstructed, as the MBs concentration decreases, especially around the cerebral cortex, the hippocampus, the thalamus, and the hypothalamus vasculature (comparing with the Allen Brain Atlas) as shown in colored dotted line in Fig.2d in 1:20 density maps.

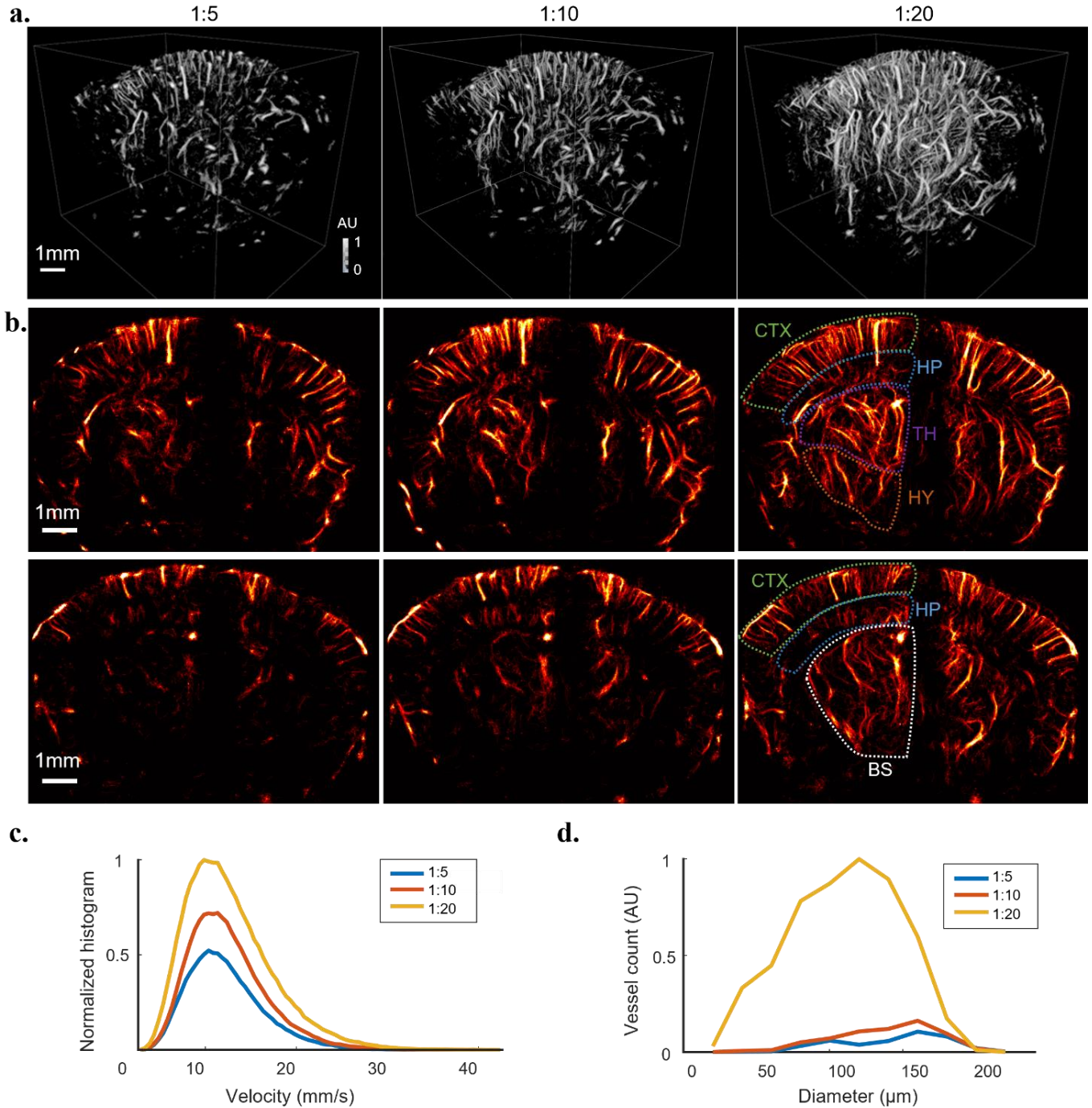

*Figure 2. Impact of microbubble's concentration on the density maps. (a) 3D volumes of density maps (for decreasing concentration of microbubbles 1:5, 1:10, and 1:20 dilution ratio (Definity: PBS) respectively from left to right. (b) Corresponding 1 mm-thick coronal slice of the brain at bregma -1.6 mm (top row) and bregma -3 mm (bottom row) for the three concentrations with main anatomical structures of the brain in dotted lines on half of the brain. (c) Impact of microbubble's concentration on the number of detected MBs and velocity distribution, (d) and vessel diameter distribution normalized by the maximum height bin value of 1:20 concentration. CTX: cerebral cortex, HP: hippocampus, TH: thalamus, HY: hypothalamus, BS: brain stem*

Fig.2c illustrates that increasing the concentration of MBs resulted in fewer detections across the entire range of imaged velocities (0 to 40 mm/s). An average loss of approximately 35% and 55% of pixel information for concentrations of 1:10 and 1:5, respectively, was observed when compared against a concentration of 1:20. Indeed, more vessels are observed at lower concentrations (Fig.2d). The total number of reconstructed segments of vessel (longer than 10 voxels, i.e. 100 μm) above 50 μm diameter is 375, 626 and 5058 for the dilution 1:5, 1:10 and 1:20 respectively.

### B. 3D Static High Resolution Brain Vasculature

Transcranial ULM using a RCA provided high resolution volumes of 11 mm × 8 mm × 8 mm of the disease model mouse brain for WT (n=3) and R6/2 (n=3) (Fig.3a). 3D renderings of different views show the main anatomical structures in the reconstructed field of view (cerebral cortex, hippocampus, thalamus, and midbrain) and the complexity of their microvasculature for all mice. The FSC gave a mean global resolution of 60±8 μm (i.e. ~$\lambda/2$) and where the PSF size in the middle of the field of view is approximately 290 μm (i.e. ~$2\lambda$) at -6 dB in lateral direction and 300 μm in axial direction. Thus, we have a global resolution of approximately one fifth of the PSF size. A 2D density slice with a 1-mm thickness of mouse #6 is shown in Fig.3b with a 10 μm pixel size. Two profile plots in the cortex reveal peaks with full width at half maximum values ranging from 19 μm to 73 μm in these regions. Velocity distribution in the whole brain of each group of mice are shown in Fig.3c with the respective peak velocity 9.75 mm/s for WT mice and 9.25 mm/s for R6/2 mice. We imaged a mean velocity of 13.1±0.4 mm/s for WT mice and 12.8±0.4 mm/s for R6/2 mice. Both distributions are not significantly different at the 1% significance level according to the two-sample Kolmogorov-Smirnov test.

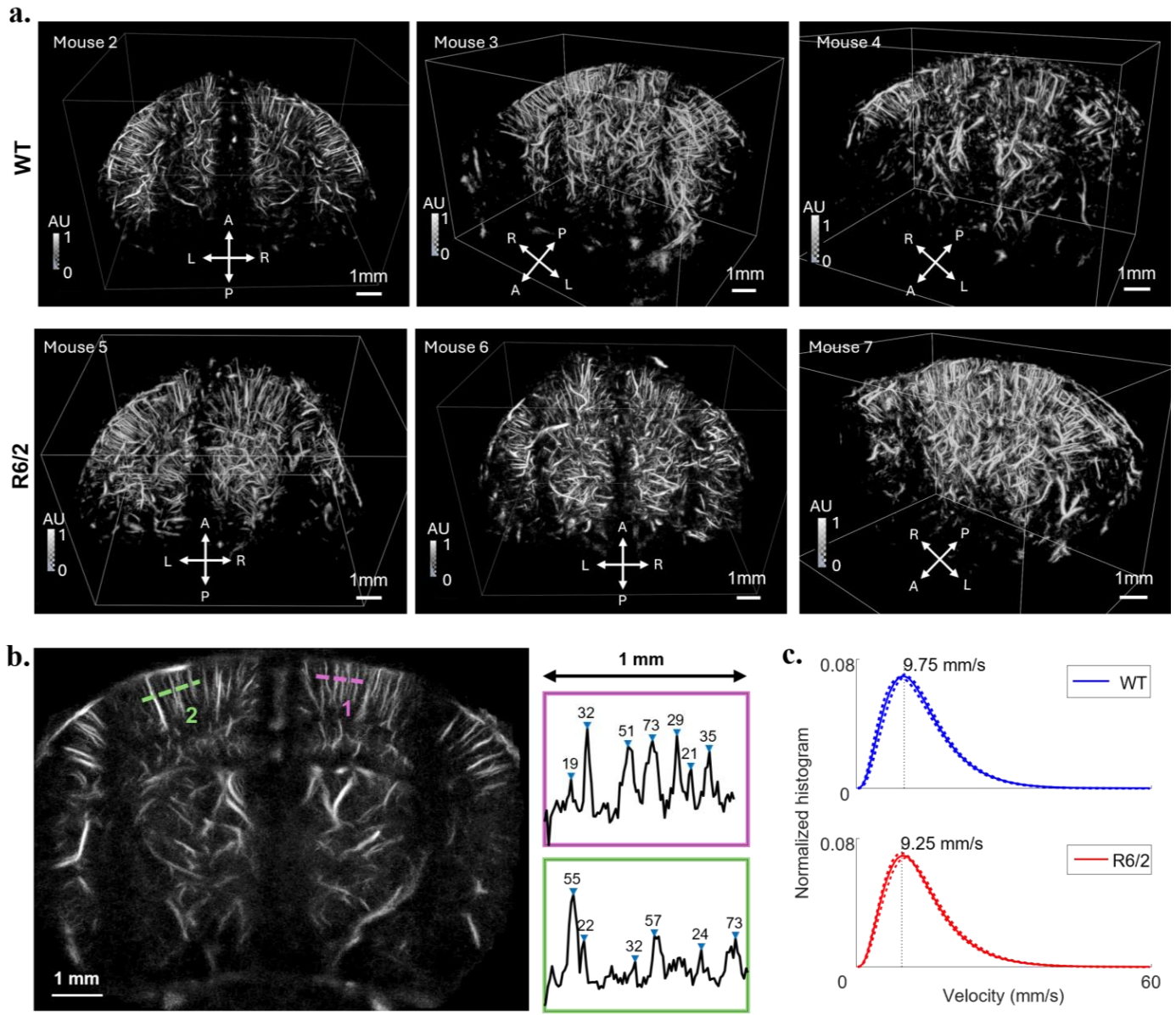

*Figure 3. (a) 3D transcranial ULM mouse brains vasculature: 3 WT (top row, mouse #2, #3 and-#4) and 3 R6/2 (bottom row, mouse #5, #6 and #7). Normalized and compressed (see Methods) density volumes using maximum intensity projection (MIP) with Amira. (b) 2D 1-mm-thickness density slice (MIP) of mouse #6 and profiles plots in the cortical region with full width half maximum values in micrometers. (c). Normalized velocity distribution (probability density function) of the whole brain of WT group and R6/2 group in solid line and the maximum and minimum velocity of the group in dotted lines.*

### C. 3D Dynamic Measurements of the Pulsatile Flow

Velocity volume of the whole brain of mouse #1 (C57BL/6J) is shown in Fig.4 and taken at time point (t = 80ms) during systole. Two zoomed-in regions of interest at different depths (cerebral cortex and brain stem) and at three different time points of a cardiac cycle show velocity (from systole t = 80ms to diastole t = 150ms) and radius variations pointed out by white arrows. A cine-loop of the dynamic velocity volume is provided in the supplementary video where two cardiac cycles can be visualized. The pulsatile pattern had a frequency of approximately 7 Hz which is consistent with the cardiac cycle of anaesthetized mice.

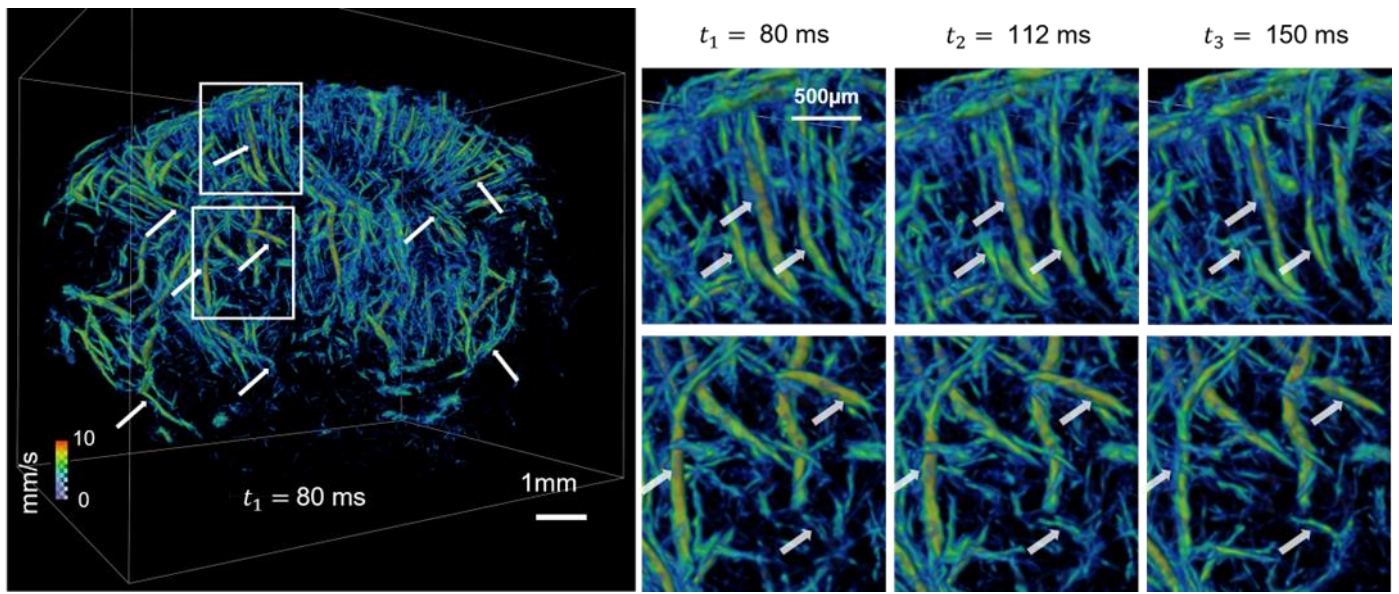

*Figure 4. 3D dynamic volume mouse brain (mouse #1 concentration 1:20). 3D velocity map of the mouse brain, taken at $t_1 = 80$ ms (systole) during the acquisition. White arrows point out vessels at different depths and areas of the brain showing velocity and radius variations along cardiac cycle ( $t_3 = 150$ ms (diastole)). See supplementary video 1 for the entire cine-loop for the whole brain.*

*D. Pulsatility measurements in cortical veins and arteries are significantly different in all animals.*

Oriented MB density maps along with the segmented cortical veins (flowing upwards in blue) and cortical arteries (flowing downwards in red) in mouse #2 (WT) are shown in Fig.5a in a coronal view. Fig.5b shows velocity waveforms of 3 cardiac cycles in cortical vessels from Fig.5a, higher velocity variations in arteries (from 10 mm/s to 20 mm/s) compared to veins (from 11 mm/s to 15 mm/s) are observed and showing similar amplitude variations between symmetrical vessels (left/right hemispheres). Pulsatility indices are shown (Fig.5c) in segmented cortical arteries and veins for all mice. A significatively higher pulsatility is observed in cortical arteries when compared to cortical veins for all mice in both groups WT and R6/2.

IV. DISCUSSION

In this study, we developed transcranial 3D Dynamic Ultrasound Localization Microscopy using an RCA to retrieve dynamic information in small vessels of the mouse brain. After fine-tuning DULM sequences and MB concentrations in the context of RCA imaging, an *in vivo* transcranial study on several mice was conducted and yielded imaging volumes of blood velocity variations over time throughout the cardiac cycles and to the measurement of pulsatility indices throughout the brain.

The impact of MB concentration was studied for bolus injection. Using 1:20 dilution provided generally better resolved volumes than higher concentration (Fig.2). Fewer MBs were detected with 1:5 and 1:10 dilutions in the whole range of velocities (up to 40 mm/s) and vessel diameter (from 50 µm to 200 µm) in the whole brain. This is most likely due to overlapped microbubbles signals [50] that have lower correlation with a PSF and increased localization error. Since DULM requires MB detections at each time point of the cardiac cycle, lower concentration further may increase acquisition time.

Transcranial ULM provided large volumes of the brain vascular images (around 80 λ × 60λ × 60λ) obtained with the optimized concentration and sequence parameters showing the main vascular structures in several WT and R6/2 mice (Fig.3), resolving vessels down to 20 µm in the cortex. Velocity distributions in the whole brain within the groups were similar, and no significant difference was observed between groups, indicating reproducibility of the method. Note that differences were not expected between models as we used pre-symptomatic young R6/2 mice [43].

Transcranial DULM provided 3D cine-loops of MBs flow in the brain across several cardiac cycles. (See supplementary materials and Fig.4). We observed a pulsatile pattern corresponding to the cardiac cycle and obtained pulsatility indices measurements from velocity changes across various vessel sizes and depths. Cortex arteries had a significantly higher PI than veins (Fig 5), consistent with prior studies [23], [51]. For this feasibility study, PI measurements were carried out in young adult mice (4 to 6 weeks) and no statistically significant difference was found between WT and R6/2. Further investigation for this R6/2 model with larger cohorts and different pathology states are the object of future work. This reproducibility indicates that 3D transcranial DULM would allow seamless integration into longitudinal studies, assessing vascular pathology progression like HD or treatment responses without susceptibility to infection or cranial window failure, although aberration correction might be necessary for older animals.

Using a RCA reduces the number of elements but only one way focusing is performed: transmit focusing is performed through compounding of multiple angles in one direction and receive focusing is done in the orthogonal direction through beamforming. This negatively affects both resolution and contrast, compared to fully addressed matrix arrays at the same frequency. Even though we tried to compensate with more angles in emission, it leads to a trade-off with respect to the achievable volume rate. Indeed, compounding a large number of emissions increases the probability of incoherent compounding, reduces the capability of the system to properly separate MBs signal from tissue signal, and the limited volume-rate degrades tracking performances. Increasing the effective volume rate could be performed as in [42] with a sliding compounding approach, which decreased the number of false detection and improved the temporal resolution. Moreover, ground truth for brain wide velocity fluctuations is currently not available at the spatiotemporal resolution reported herein. To address this, considering the coupling of 3D DULM with optical techniques like OCT or TPM [7], [9] might offer a way forward. This combination could validate both the anatomical structure and pulsatility measurements in superficial vessels. Furthermore, bolus injections induce MB concentration variation during the acquisition time. Continuous infusion may reduce this variation and provide more stable measurements as well as shorter acquisition time but remains difficult to implement in practice due to the small injection volumes in mice. Experiments were carried out on a small cohort of young mice (n=6). Larger cohorts with different stages of the HD pathology are needed to provide more accurate statistical evaluations.

This study's outcomes illuminate several avenues for future research and application. By understanding the impact of pulsatility on the brain, it could help monitor treatments and, for instance, delve deeper into the influence of the brain microvasculature in diseased models like the R6/2 mice. Focusing on pathological models offers a way to study the influence of pulsatility across the entire brain. Since 3D DULM requires higher concentrations of MBs without increasing the acquisition time, challenges in localization and tracking present another frontier. We could address these issues using deep learning approaches [52], [53] or by tracking in the spatiotemporal domain prior to MB localization at a subwavelength resolution [54]. Other beamforming approaches could be performed like XDoppler [37] or frame multiply and sum [55] to reduce sidelobes level and false detections. Finally, some regions in the brain remained noisy and difficult to image due to skull aberrations, the presence of ventricles, or shadowing from bigger vessels. In this study, we did not apply any aberration correction but the generalization to RCA probe of recent work on aberration correction [22], [56], [57], [58], [59] could improve the image quality and resolution of 3D DULM. Lastly, translating this method for human application necessitates a transition to lower frequency probes in order to image through the thicker human skull with a broader field of view, but compromises the volume rate. Nevertheless, the potential for gaining insights into human brain hemodynamics makes this a worthwhile pursuit.

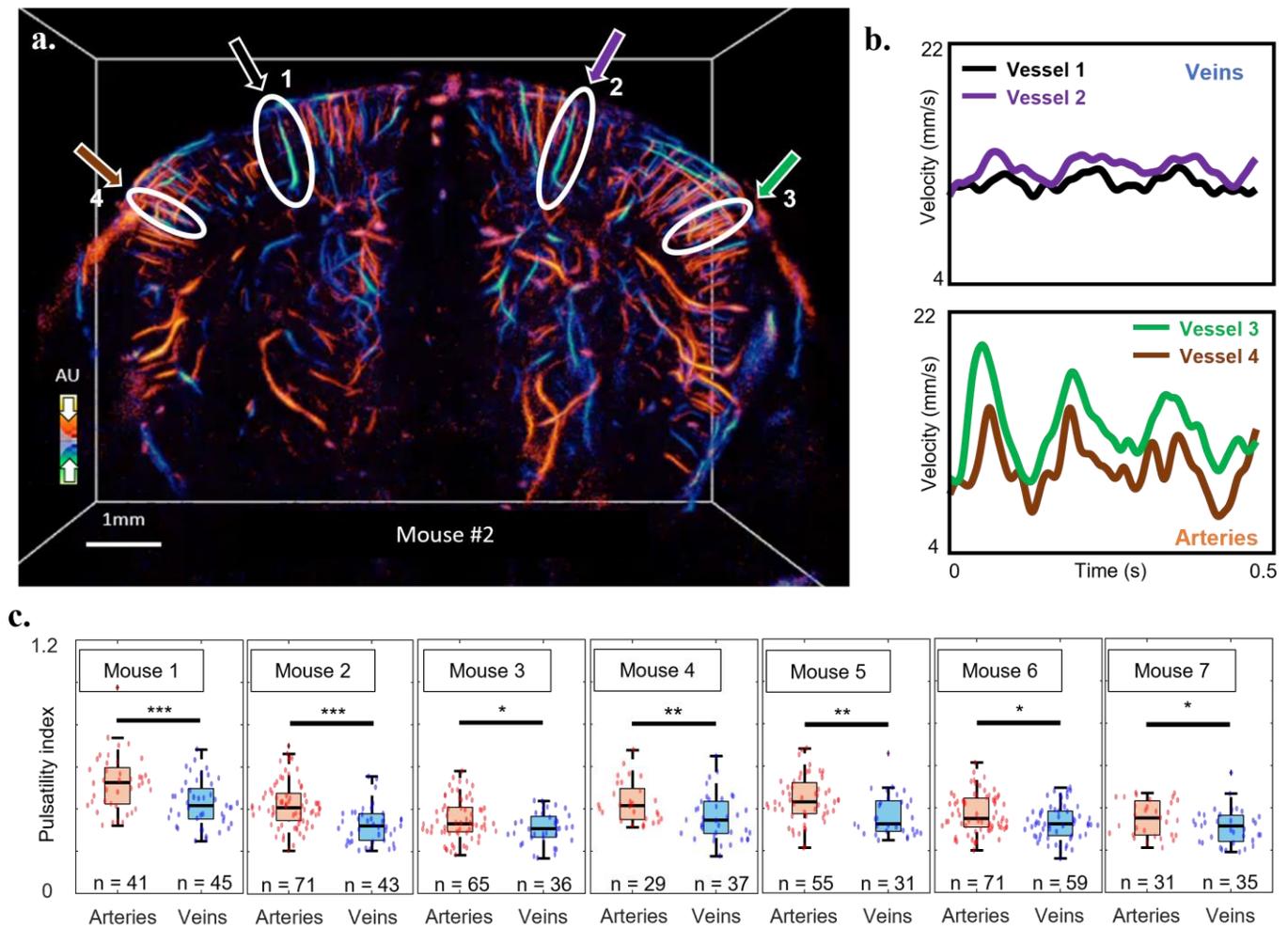

*Figure 5. DULM-based PI could differentiate veins from arteries in the mouse cortex. (a) Coronal view of oriented microbubbles (MB) density volumes (red: MBs flowing downwards, blue: MBs flowing upwards). (b) Pulsatile pattern of veins and arteries respectively selected from (a). (c) PI calculated for each cortical segmented arteries and veins for n=7 mice with their respective number of vessels kept for PI measurement. Levels of significance are given as \*: p< 0.05, \*\*: p<0.01, and \*\*\*: p< 0.001.*

V. CONCLUSION

We demonstrated the feasibility and reproducibility of transcranial 3D DULM using a RCA probe in the mouse brain, confirming its robustness of performing deep and non-invasive pulsatility measurements considering the 3D structure of the brain. We were able to generate dynamic volume of MBs flowing in the vasculature and to extract pulsatility measurements from the mouse cortical vessels but also from deeper vessels in the brain. This approach provides promising results and could lead to a better understanding of the pulsatility impact on cognitive decline and other pathological models. 3D DULM could become a non-invasive and portable tool to monitor brain blood flow to assess the anatomy and function of the brain microvasculature.